# Transient vortex dynamics and evolution of Bose metal from a 2D superconductor on MoS$_2$

Sreevidya Narayanan[1], Anoop Kamalasanan[1], Annu Anns Sunny and Madhu Thalakulam[*]

Indian Institute of Science Education & Research Thiruvananthapuram, Kerala 695551, India

## Abstract

The true character of physical phenomena is thought to be reinforced as the system becomes disorder-free. In contrast, the two-dimensional (2D) superconductor is predicted to turn fragile and resistive away from the limit $I \to 0$, $B \to 0$, in the pinning-free regime. It is intriguing to note that the very vortices responsible for achieving superconductivity by pairing, condensation, and, thereby reducing the classical dissipation, render the state resistive driven by quantum fluctuations in the $T \to 0$. While cleaner systems are being explored for technological improvements, the 2D superconductor turning resistive when influenced by weak electric and magnetic fields has profound consequences for quantum technologies. A metallic ground state in 2D is beyond the consensus of both Bosonic and Fermionic systems, and its origin and nature warrant a comprehensive theoretical understanding supplemented by in-depth experiments. A real-time observation of the influence of vortex dynamics on transport properties so far has been elusive. We explore the nature and fate of a low-viscous, clean, 2D superconducting state formed on an ionic-liquid gated few-layered MoS$_2$ sample. The vortex-core being dissipative, the elastic depinning, intervortex interaction, and the subsequent dynamics of the vortex-lattice cause the system to behave like an overdamped harmonic oscillator, leaving transient signatures in the transport characteristics. The temperature and magnetic field dependence of the transient nature and the noise characteristics of the magnetoresistance confirm that quantum fluctuations are solely responsible for the Bose metal state and the fragility of the superconducting state.

[1] Equal contribution

[*] madhu@iisertvm.ac.in

**Introduction**

Quantum technology, though exploits systems in reduced dimensions, deals with both quantum-confined and macroscopic quantum states. The signatures of confinement-driven quantum systems, viz., discreteness and quantization, are reasonably well understood; nevertheless, the behaviour of macroscopic and many-body quantum states, such as superconductivity, ferromagnetism and fractional quantum Hall states, in reduced dimensions can be non-trivial. Superconductivity in three-dimension (3D) is a state with zero electrical resistance at non-zero temperatures. When the superconducting coherence length $\xi$ exceeds the sample thickness, the system is considered to be in the two-dimensional (2D) limit[1]. Spontaneous symmetry breaking in reduced dimensions, though forbidden by the Mermin-Wagner theorem[2], the 2D superconducting transition can be understood in terms of the Berezinskii-Kosterlitz-Thouless (BKT) transition, where the generation and pairing of vortices along with the condensation of Cooper pairs pave the way to the zero-resistance state[3,4]. Vortices are topological excitations in the superfluid, generated by thermal or quantum fluctuations. The zero-resistance state is a virtue of the unlimited superfluid phase coherence. Factors disrupting the order parameter, such as disorders and vortex motion, contribute to dissipation and are manifested as resistance in transport measurements. Arresting vortex motion by pinning at the surface or by the interaction between the vortex lines enables the zero-resistance state in type-II 3D superconductors[5,6]. Historically, 2D superconductivity has been hosted by thin-film[7] or interfacial superconductors[8] which are inherently disordered and on which the vortices have limited mobility. The vortex velocity varies inversely with the superfluid viscous coefficient. Clean 2D systems in the pinning-free limit possess a very low superfluid viscous coefficient, consequently, prone to the generation and motion of vortices by external driving currents and magnetic fields, making the superconducting state fragile[9,10]. As a result, the zero-resistance state is guaranteed only in the limit of $I \to 0$, $B \to 0$[10,11].

Fundamentally, understanding the behaviour of vortices in the substrate-free limit is a topic of great interest not just to the field of superconductivity but also to other superfluid systems[12,13]. Quantum technologies such as quantum information processors and quantum electrical metrology circuits rely on planar architecture. While the semiconductor-based technologies are already hosted by 2D electron systems[14], the superconducting circuits are rapidly approaching the 2D limit and 2D layered systems[15,16]. The vortex dynamics in these systems will have a profound effect on the design and performance of these devices[17]. While the application of static and time-varying electric and magnetic fields is unavoidable, cleaner systems transform into the BM state and turn dissipative under the application of external fields. In the substrate-free regime, vortex motion is predominantly subsided by intervortex interaction[6] and the subsequent formation of the vortex lattice once a sufficient number of vortices are injected into the system by an external field[18,19]. The dissipative core makes the vortex motion overdamped[6,20], and there should be a timescale associated with the formation of the vortex lattice and the subsequent reaching of the low-resistance equilibrium state. This time-scale need to depend on the superfluid viscosity, vortex mobility, intervortex interaction, vortex lattice elasticity, and pinning strength. The temporal behaviour of electrical transport against different regimes of vortex dynamics is a much wanted figure-of-merit for device technology, which forms the motivation of the work presented in this manuscript.

Systems, owing to the dynamics of vortices, can exhibit a spectrum of non-trivial intermediate quantum states between the superconducting and the insulating states[21–23]. In this regard, on clean crystalline van der Waals (vW) 2D superconducting systems with substantially low vortex pinning, evolution of two types of dissipative quantum metallic (QM) states, viz., the quantum tunnelling of vortices and the Bose metal (BM) states[24], uncharacteristic for Bosonic systems. The BM state has been visualized as a state with incoherent motion of uncondensed vortices and Cooper pairs resulting from the dynamical gauge field fluctuations and the retardation effects[25,26]. It has also been argued that the quantum phase fluctuations create a glassy phase over which the superfluid motion is dissipative[27,28]. A clear consensus on the exact origin and nature of the BM states is still lacking, largely due to the dearth of experiments in this regime[29–31]. A cross-over from the 2D superconducting to the BM phase at very low temperatures and magnetic fields has been observed on ultra-clean crystal of bilayer $NbSe_2$[29] and few-layered $1T-MoS_2$[30]. Though these systems, such as $NbSe_2$, $TaS_2$ and $1T-MoS_2$, are cleaner and possess a lower disorder density compared to thin film and interfacial materials, they are highly unstable against ambient conditions and standard sample fabrication processes, limiting the system from reaching the pinning-free limit[30,32,33]. In contrast, on semiconducting transition metal dichalcogenides such as $MoS_2$ and $WS_2$, electrostatic doping using ionic liquid (IL) can induce carrier densities $\sim 10^{14}/cm^2$, to observe 2D superconductivity[34–37].

Here we explore the nature and fate of a 2D superconducting state in the disorder-free limit on a few-layered $MoS_2$ flake exploiting electrolytic gating technique using DEME-TFSI (N,N-diethyl-N-methyl-N- (2-methoxyethyl) ammonium bis (trifluoromethylsulfonyl)- imide) as the IL. The device shows metallic behaviour down to a temperature ~ 4 K and transforms to the superconducting state with further reduction in temperature. The 2D nature of superconducting state is confirmed by the presence of BKT transition and magnetotransport measurements. The magnetotransport measurements also show that our sample hosts a highly fragile 2D superconducting state with a very low superfluid viscous coefficient and high vortex mobility. Consequently, we observe the emergence of the BM state under the application of a very low magnetic field. Vortex motion leaves transient signatures in the magnetoresistance. By analysing the temporal behaviour of the resistance, we establish a one-to-one correlation between the vortex dynamics and the quantum phases realized on our system against the magnetic field, the driving current, and the temperature. Our attempts on understanding the behaviour of a true 2D superconducting state under the application of external fields and currents carry a lot of importance for the future quantum technology and also shine light on the behaviour of 2D superfluid systems under the action of external driving forces.

**Results & Discussion**

**Phase diagram of disorder-free 2D superconductor**

Fig. 1(a) shows a perspective optical image of a few-layered MoS$_2$ device similar to the measured one with the IL droplet covering the sample and the liquid-gate electrode. The inset to Fig. 1(a) shows an optical image of the device. From the AFM height profile (not shown), we extract a thickness of ~ 9.53 nm, which corresponds to ~15 layers of MoS$_2$. All the data presented in this manuscript are acquired with a liquid-gate voltage, $V_{IL}$ = 1.25 V. After dropping the IL and applying 1.25 V on to the liquid-gate the four-probe (4P) resistance of the sample reduces to 49 Ω at 150 K, well below the freezing point of the IL. We find that the resistance continuously drops as the sample is cooled down from 150 K and plateaus at ≈ 6.8 Ω for temperatures between 10 K and 4 K indicating a metallic nature of the sample.

Fig. 1(b) shows the 4P resistance of the sample as a function of temperature between 4 K and 10 mK. We extract a carrier concentration of $9.2 \times 10^{14} cm^{-2}$ from the low-field Hall measurement data acquired at 4K, shown in the Supporting Information SI-1. From the temperature dependence of resistance and the Ioffe-Regal criteria, [Supporting Information SI-2] we confirm the metallic nature of the sample. From Fig. 1(b) we observe that the resistance starts dropping rapidly below ~ 3 K and reduces to an immeasurably low value at ~1.15 K, indicating a transition from the metallic to the superconducting phase. The critical temperature

$T_c$, the temperature at which resistance drops to 90% of the normal state value, ≈ 2.88 K for our sample.

In IL gating, by virtue of the extremely high electric field created (~10 MV/cm)[38] at the channel by the accumulated ions, the electrons are held mostly in the vicinity of the topmost layer, and the device is considered to be in the monolayer limit[39] and expected to host a 2D superconducting state. We note here that the normal state resistivity of the sample in the entire temperature range is significantly below the upper limit, $h/(4e^2) \sim 6.45$ kΩ to realize 2D superconductivity[24].

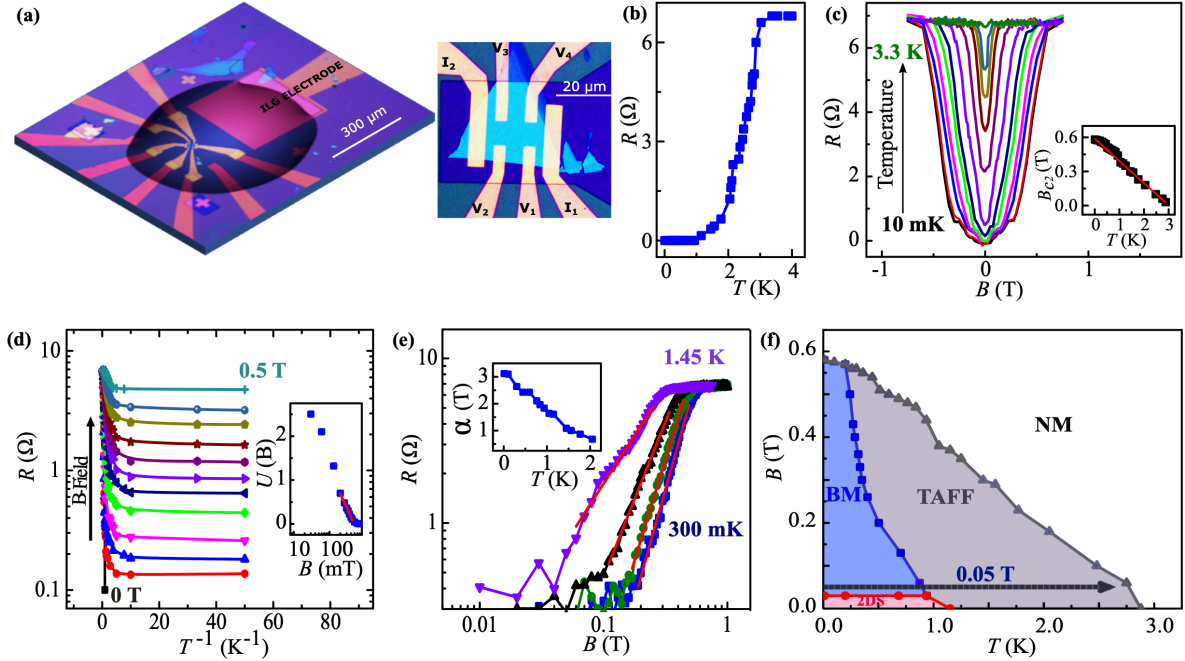

Fig. 1. 2D superconductivity & Bose metal states on MoS$_2$. (a) Perspective optical image of the liquid gated MoS$_2$ FET (left) and an optical image of the MoS$_2$ flake (right). Current is sourced across the probes I$_1$ and I$_2$. Longitudinal Resistance ($R_{XX}$) is measured across the probes V$_1$ and V$_2$ and transverse Resistance ($R_{XY}$) is measured across the probes V$_1$ and V$_4$. Ionic liquid gate electrode is marked. (b) Plot of the resistance versus temperature at 0 T showing the transition to the superconducting state. (c) The Magnetoresistance of the device for temperatures ranging from 10 mK to 3.3 K. Inset: critical field versus temperature. The red-straight line is a fit to the 2D Tinkham model. (d) Arrhenius plot of the longitudinal sheet resistance of the device for different perpendicular magnetic fields. Inset: semi-log plot of the activation energy, $U(B)$ extracted from the main panel. The red solid line is a fit to the data using the equation $U(B) = U_0 \ln(B_0/B)$. (e) Magnetoresistance for various temperature in the log-log scale. Red solid lines are fits to the power law $R \sim (B - B_C)^{\alpha(T)}$. Inset: α versus T. (f) $B - T$ phase diagram. Boundary between the normal metal (NM) and TAFF phases are represented by the grey triangles. The red circles show the boundary of the 2D Superconductor. The blue shaded region shows the Bose metal phase.

Fig. 1(c) shows a few representative magnetoresistance plots for perpendicular magnetic field orientation, acquired using the stabilized mode, for temperatures ranging from 10 mK to 3.3 K. The sample makes a transition from the superconducting to the normal state progressively as the magnetic field is increased. We regard the field at which the resistance assumes 90% of the normal state value as the upper critical field, $B_{C2}$ which is 0.58 T at 10 mK. The inset to Fig. 1(c) shows a plot of $B_{C2}$ versus $T_C$ extracted from the magnetotransport measurements. According to the 2D Tinkham model, for perpendicular field orientation, the variation of upper critical field with temperature is given by $B_{C2} = \frac{\phi_0}{2\pi \xi_{GL}(0)^2}\left(1 - \frac{T}{T_C}\right)$ where, $\phi_0$ is the flux quantum, $\xi_{GL}(0)$ is the Ginzburg – Landau coherence length[1]. The red line is a fit to the data using the Tinkham model, from which we extract, $\xi_{GL}(0) = 23.76 \pm 0.48$ nm. The excellent agreement with the Tinkham model confirm that our sample hosts a 2D superconducting state. Moreover, the BCS coherence length, $1.35\,\xi_{GL}(0) = 32$ nm $< l_e(\text{the mean} - \text{free path}) = 67$ nm points to the fact that our device operates in the clean 2D superconducting regime[29]. The residual resistance ratio ($RRR$), the ratio of the resistance at high temperature (150 K in our case) to that just above $T_C$ is a measure of the crystallinity of the sample. For our sample, $RRR = R(150\ K)/R(4\ K) \sim 7.2$, which is comparable to other reported clean 2D systems and considerably higher than that for disordered systems[33,40].

Clean 2D superconducting systems transform into a dissipative QM state under the application of perpendicular magnetic fields[24,29]. Fig. 1(d) shows an Arrhenius plot of the sheet-resistance for different values of perpendicular magnetic fields. At lower temperature ranges the resistance saturates to low values and shows an activated behaviour with field, a landmark signature of QM state[29]. In this regime, the resistance goes to zero when $B \rightarrow 0$, suggesting that the phase coherence is limited by quantum fluctuations not by temperature. In addition, the device shows an activated behaviour at higher temperatures. The inset to Fig. 1(d) shows a plot of the activation energy $U(B)$ versus $B$-field extracted from the activated regime of the traces shown in Fig. 1(d), a fit to which using the equation[29] $U(B) = U_0 \ln(B_0/B)$ yields $U_0 = 1.68$ K and $B_0 \sim 0.56$ T. The activation energy becomes almost zero around $B_0 \sim 0.56$ T, allowing the free-flow of vortices above this field. We note here that the $B_0$ we obtain lie very close to the upper critical field $B_{C2} = 0.58$ T.

An exponential dependence of resistance to the field manifests the quantum creep of vortices[41] while a power law dependence[25,29], $R \sim (B - B_{C0})^{\alpha(T)}$, is a signature of the BM state, where $B_{C0}$ is the critical field for the superconductor to BM transition. Fig. 1(e) shows the magnetoresistance for different temperature in the log-log scale, and the linear scaling observed indicates a power-law dependence on the field, confirming the emergence of a BM phase on our sample[25,29]. The inset to Fig. 1(e) shows the extracted exponent $\alpha$ as a function of temperature consistent with other studies on the BM phase[29,30]. We also tried fitting our data with the quantum-creep model[41], though obtained poor results, negating the presence of quantum creep in our sample [Supporting Information SI-3].

Fig. 1(f) shows the magnetic field-temperature $(B - T)$ phase diagram for the 2D superconductor extracted from the magnetotransport measurements. The boundary between the 2D superconducting phase (2DS) and the BM is assumed to be the point where the system makes a transition to a measurable resistance. The boundary between the BM and thermally activated flux flow[29,42] (TAFF) regimes, the blue squares, is marked by the onset of thermally activated transport behaviour, $R \sim \exp(U(B)/T)$ from the saturation regime $R = R(B)$ in Fig. 1(d). The boundary between the TAFF and the normal state (grey triangles) are extracted from the critical fields obtained from Fig. 1(c).

The 2D superconducting phase in Fig. 1(f) confines to a very narrow magnetic field regime ( ≲30 mT) unlike that for systems reported previously by us[30] and also by others[43], a signature of a clean 2D superconductor with negligible pinning. In the absence of pinning forces, there will be a voltage caused by the flux flow at all temperatures and applied currents due to quantum fluctuations[9]. As a result, the resistance shows a smooth and continuous rise from the zero value to small non-zero values with the field. Assuming all carriers are confined to the topmost layer, we consider a superconducting thickness ~1nm and estimate an upper bound for the superfluid viscous coefficient $\eta_0 \cong \frac{\varphi_0^2}{2\pi\xi^2 \rho_n} \sim 10^{-8}$ kg m$^{-1}$s$^{-1}$ that is comparable to that observed on other systems with negligible pinning[44]. $\rho_n$ is the normal-state resistivity.

**Dissipative transport due to current induced vortex generation and depinning**

Fig. 2(a) shows $I - V$ characteristics acquired at various perpendicular magnetic fields. As the magnetic field is increased, the extent of the flat region reduces, and eventually, the $I - V$ traces turn into those of a normal metal. For $B = 0$ T, the traces obtained with the forward (black) and reverse (red) sweep directions is void of any hysteresis, suggesting negligible pinning; the pinning-depinning mechanism causes hysteresis in the $I - V$ characteristics in the vicinity of critical current[43,45]. We observe the development of a voltage across the sample even at driving currents as low as ~ 8 µA. The current-voltage relation in the limit of negligible pinning is given by $V = C \left(\frac{\varphi_0^2 L}{\eta_0 S}\right) I^\alpha$ where L is the distance between the voltage contacts, S is the cross-sectional area of the current flow and, $\alpha$ and $C$ are temperature-dependant constants determined from the fit to the $I-V$[40] shown in the right inset to Fig. 2(a). The number density of free vortices induced by the applied current $n_f = CI^{\alpha-1}$. The vortex velocity $v_f = \frac{\left(\frac{V}{L}\right)}{CI^{\alpha-1}\varphi_0}$. The left-inset shows the calculated $n_f$ and $v_f$. For a given device resistance, we compare the $n_f$ estimated from the $I-V$ characteristics shown in Fig. 2(a) to that from the magnetotransport data shown in Fig. 1(c) and find a very good agreement between the measurements. For instance, $R\,(95\,\mu A) = R\,(0.5\,T)$, accordingly, we obtain $n_f = 1.15 \times 10^{14}\,m^{-2}$ for $I = 95$ μA while $n_f = 2.4 \times 10^{14}\,m^{-2}$ for $B = 0.5$ T, suggesting the rise in the device resistance in the $I-V$ characteristics is due to dynamics of current-induced vortices. The calculated vortex velocity, shown in blue in the upper-left inset to Fig. 2(a), agrees very well with that observed in a similar low viscous superconducting system[44]. The differential resistance shown in Fig. 2(b), for $B = 0$ T, exhibits discrete jumps followed by plateaus in the conductance, and this behaviour has been attributed to the formation of vortex channels[19,44,46–48].

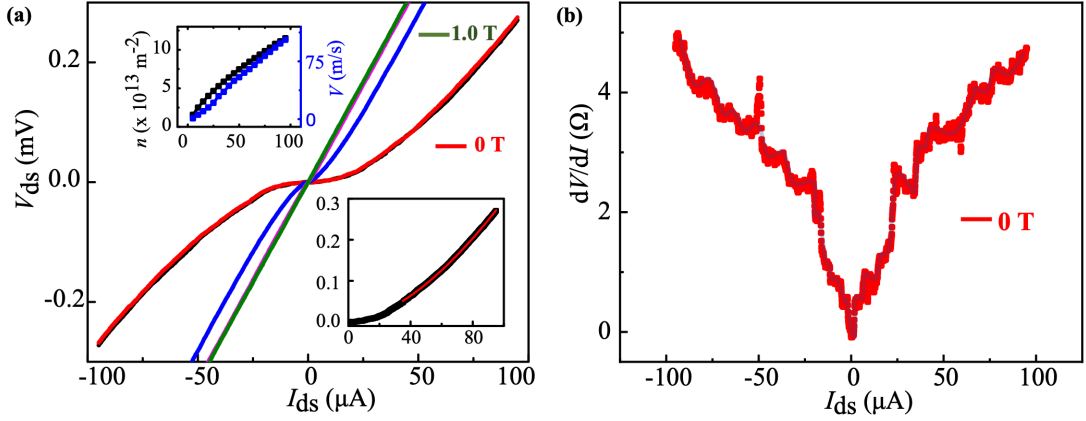

**Fig. 2. $I-V$ characteristics.** (a) $I-V$ characteristics at 10 mK for different perpendicular fields; 0 T (black-forward & red-reverse), 0.25 T (blue), 0.5 T (magenta) and 1 T (green). Bottom inset shows the power law fit to $I-V$ curve in the current range of 35 μA to 100 μA. Top inset shows the calculated vortex number density (black) and velocity of vortices (blue) versus applied current at 10 mK. (b) $dV/dI$ plots at 10 mK for $B = 0$ T. Grey line is a guide to the eye marking the steps in $dV/dI$ plot.

**Vortex dynamics in the substrate-free limit results transient transport characteristics**

The introduction of a magnetic field will populate the sample with mobile vortices, disrupting the phase coherence. The pinning strength, intervortex interaction, driving force, and quantum fluctuations collectively decide the ground state of the system. The vortex-vortex interaction depends on the gradient of the magnetic field due to themselves[6], which is substantial at London penetration length scales. At very low fields, the vortices are highly mobile owing to negligible intervortex interaction[5]. Once the inter-vortex distance approaches the London penetration length, the vortex-vortex interaction will dominate the substrate-vortex interaction, resulting in the formation of a triangular-like spatial ordering of the vortices in 2D[6]. This will limit individual vortex motion and dissipation. At higher fields, the inter-vortex separation, the size of the vortex, and the field gradient between vortices diminish[6], making the interaction weak and causing a melting of the vortex lattice. The vortex lattice is found only in the intermediate fields[5]. Since the resistance of the system reflects the vortex dynamics, the application of a magnetic field and the subsequent vortex rearrangement cause a transient behaviour in the device resistance in the pinning-free limit. We search for signatures of this transient behaviour as we take the system across the 2D superconducting to BM and TAFF regimes.

Fig. 3(a) shows magnetotransport measurements conducted at 10 mK. The magnetic field is swept from +1 T to -1 T at a rate of 0.1 T/min in the continuous sweep mode. As a guide to the eye, the data acquired using the stabilized sweep mode is shown in black. The inset shows the magnetoresistance for the forward sweep direction. For both sweep directions, we observe a sharp rise in the resistance when the system traverses through the zero-field value, followed by a slow partial drop, creating a secondary minimum. Thereafter, the resistance follows that of the stabilized sweep. Fig. 3(b) shows the results of similar measurements conducted for various field sweep rates. We find in both Fig. 3(a) & (b) that the traces in the forward and reverse directions are perfect mirror images of each other about the $B = 0$ plane. The transient nature of the magnetoresistance is observed in the vicinity of zero field only when the polarity of the field is switched. In all other regimes, the magnetoresistance obtained using the stabilized and continuous modes agree with each other. When the field switches direction from the zero field value, an avalanche of vortices is formed in the system making it disordered, resulting in a sudden rise in the resistance. As the field is increased further, the number of vortices and thus the intervortex interaction increase, helping to regain phase coherence and lowering the resistance. From Fig. 3(b), we see that both the rise and the secondary minima also increase with the sweep rate. When the field sweep rate is increased, the vortices are introduced at faster rates, and the system does not get sufficient time to relax and regain phase coherence, causing the rise and the secondary minima to move to higher values.

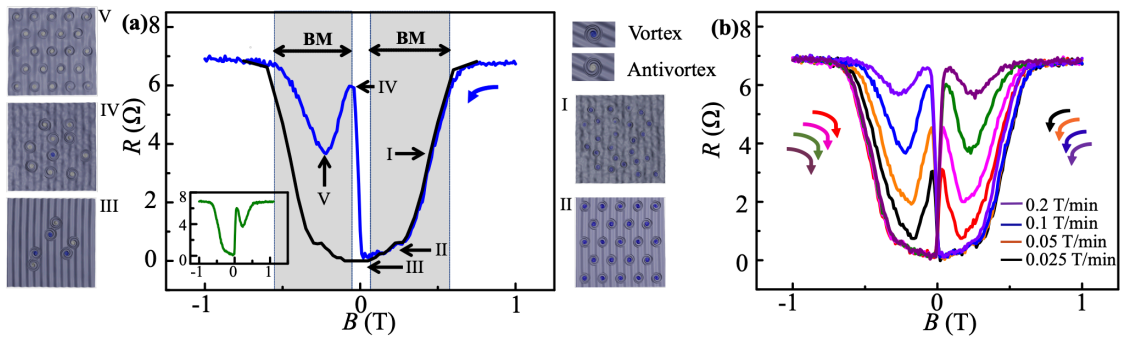

**Fig.3. Magnetoresistance transient behaviour**. **(a)** Blue-trace: Magnetoresistance for +1.0 T to − 1.0 T sweep direction, the blue arrow shows the direction of the field-sweep. The shaded regions represent the BM state. Illustrations I through V represent various regimes of vortex dynamics. Inset shows the magnetoresistance for −1.0 T to +1.0 T sweep direction. Magnetoresistance for the stabilized sweep mode is shown in black. **(b)** Magnetoresistance for perpendicular magnetic fields for different field sweep rates at 10 mK in the forward and reverse field-sweep directions. The Arrows represent the direction of the field sweep.

The illustrations indexed I through V in Fig. 3(a) show various regimes of vortex dynamics the system traverses as the B-field is varied from 1 T to -1 T. We note from the phase diagram that at this temperature, the BM state extends up to ~ 0.58 T and for higher fields the system goes directly to normal metal state. The shaded regions in Fig. 3(a) represent the extend of the BM state, extracted from the phase diagram shown in Fig. 1(f). Point I corresponds to the BM state consisting of uncondensed Cooper pairs and vortices. As the field is lowered, the vortex lattice structure is progressively built[18], taking the system to a lower resistance regime (point II). Point III corresponds to the 2D superconducting state where the systems consist of isolated condensed vortex-antivortex pairs and Cooper pairs. Once we crossover to the opposite field direction, the vortex avalanche causes the rise in the resistance (point IV). At point V, the intervortex interaction helps to regain phase coherence partially, causing the resistance to drop. Since the field is continuously swept and the system is not given sufficient time to relax to the ordered state, the resistance at point V does not attain the minimum value (~ 0.45 Ω) observed in the stabilized sweep. We note here that the data points on the black trace shown in 3A are acquired using the stabilized sweep mode and will not show the above mentioned behaviour.

From Fig. 3(a) & (b), we observe that the resistance spikes until a field of ~50 mT and droops upon any further increase of the field. In addition, from the phase diagram shown in Fig. 1(f), we note that a field increment of 50 mT from 0 T takes the system from the 2D superconducting to the BM state unambiguously. To understand the nature of the 2D superconducting to BM transition and capture the timescale of the vortex dynamics, we inspect the temporal behaviour of the resistance while we take the system from the 2D superconducting to the BM state. We record the resistance as a function of time while performing a field increment of 50 mT from the zero field value at a rate of 0.1 T/min, as shown in the lower panel of Fig. 4(a). The regions enclosed by the vertical lines in Fig. 4(a) represent the duration of the field sweep. As the field is increased, the resistance surges to ~5.54 Ω and then starts dropping as the 50 mT endpoint is reached, subsequently saturates to a very low non-zero value. The initial sharp rise is due to the sudden introduction of highly mobile vortices disrupting phase coherence[49]. The drop in resistance is due to the formation ordered arrangement of vortices limiting the vortex motion. In the clean limit, there will be elastic de-pining and collective motion of the vortex lattice[6] under any small driving force, causing the resistance to saturate to non-zero values. We define a time-constant $\tau$ ($\approx$ 38 s) as the time taken by the resistance of the system to drop down to 50% of the rise. This we identify as a measure of the duration to reorganize the vortices into an ordered array once a sufficient number of vortices are abruptly introduced into the system. As the system approaches the clean-limit, the relaxation time will be longer[49]. When we take the field from 50 mT to 100 mT [middle-panel, Fig. 4(a)] by the same ramp rate, the rise in the resistance and the corresponding time-constant are much lower, ~ 2.138 Ω and ~ 30.74 s respectively. When we take the system from 100 mT to 150 mT, [top-panel, Fig. 4(a)] we do not observe any sudden rise in the resistance. We observe a small overall change in the resistance ~ 0.2 Ω, which is the natural response of the superconducting state to the increase in the field. For 50 mT to 100 mT step, the existing vortex lattice and the inter-vortex interactions limits the vortex motion and dissipation. For the 100 mT to 150 mT field-step, the intervortex interaction is even stronger and we do not observe any noticeable sudden rise.

Resistance is caused by vortex motion induced by fluctuations. We search for signatures of these fluctuations by extracting the noise power spectral density (NPSD) from the resistance of the system, as shown in Fig. 4(b). NPSD is extracted from the tail section of the resistance versus time traces marked by the vertical dashed lines in Fig. 4(a) so that the system is in the ground state. The NPSD consists of noise from the system and the measurement circuits with a bandwidth of 1 Hz. We observe that the superconducting phase exhibits the minimum integrated noise power ($NP_{int}$). As the system is taken into the BM state by performing small field increments, the $NP_{int}$ also increases progressively, as shown in the top three panels (back, red, and blue) in Fig. 4(b) suggesting that the noise is driven by quantum fluctuations.

The nature of vortex dynamics depends on the phase the system is being transformed into, and the time constant is a measure of the duration for stabilizing this target phase. To get more insight into the nature of BM, we inspect the time constant $\tau$, described in Fig. 4(a) lower panel, for the temperature range of 10 mK through ~ 2 K, as shown in Fig. 4(c). The dashed line in Fig. 1(f) parallel to the temperature axis represents the array of target states the system goes through as the temperature is varied. One can infer from Fig. 1(f) that for a field step of 50 mT from 0 T, the target phase is a BM state until $T \sim 938$ mK and moves to the TAFF phase for higher temperatures. The initial state remains a 2D superconducting state until a temperature ~ 1.15 K. The BM state is a quantum phase where the phase fluctuations control the dynamics of the vortices. As long as the target is BM, the temperature should not play any significant role. This feature is underlined by the behaviour of $\tau$ versus $T$ shown in Fig. 4(c). In Fig. 4(c), the shaded region is a guide to the eye, corresponds to the BM state. $\tau$ is nearly unchanged (~ 38 - 40 s), except for any measurement errors, till $T$ ~ 938 mK. A further increase in the temperature takes the target state to TAFF regime, where the thermal fluctuations also starts contributing to vortex dynamics. This is evident from the increase in $\tau$ once the target phase moves beyond the 938 mK point; here, the initial phase is 2D superconducting, while the target phase is TAFF. For further increments in the temperature, the time constant also rises in accord suggesting the dynamics is dominated by the thermal fluctuations, a signature of TAFF phase.

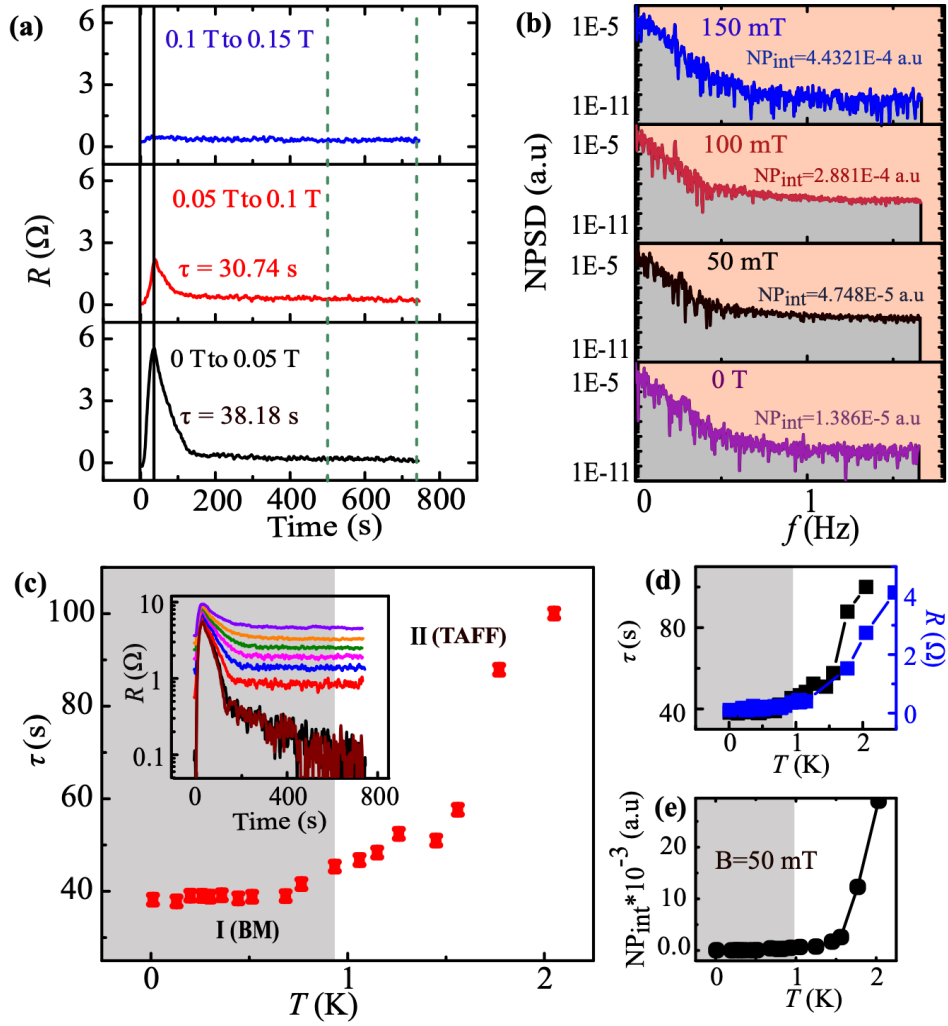

**Fig. 4. 2D superconductor − BM phase transition.** (a) Transient behaviour of resistance as the magnetic field is stepped from 0 T to 50 mT (lower), 50 mT to 100 mT (middle) and 100 mT to 150 mT (upper) at 10 mK and, $\tau$ the corresponding time-constants. The vertical lines represent the start and the end points of the 50 mT field increment. (b) Noise power spectral density extracted from the resistance versus time traces for the 2D superconducting (violet) and the BM state; black (50 mT), red (100 mT) and, green (150mmT). (c) $\tau$ versus temperature for the field step 0 T to 50 mT, the 2D superconductor to the BM transition. The shaded region represents the the BM state while region-II the TAFF phase. Inset: transient behaviour of resistance (resistance versus time) for a field step of 0 T-50 mT for various temperatures starting from10 mK (brown), 514 mK (black), 935 mK (red), 1.15 K (blue), 1.25 K (magenta), 1.45 K (green), 1.56K (orange) and, 1.77 K (violet). (d) $\tau$ versus temperature (left) and device resistance versus temperature (right). (e) Integrated noise power, $NP_{int}$ versus temperature at 50 mT field, as the system is taken from the BM to the TAFF regime.

Top-left inset to Fig. 4(c) shows the transient response of the resistance for the 0 to 50 mT field-step for a few representative temperatures starting from 10 mK through 1.77 K. Supporting Information SI-4 shows the same for the 0 to −50 mT field-step exhibiting similar behaviour. We note here that not only the time-constant but also the temporal behaviour of resistance shows a sudden variation as the target system switches from the BM to the TAFF regime. For temperatures above 938 mK, corresponding to the TAFF regime, the resistance shows an exponential-like drop with time, whereas for all temperatures below 938 mK, the resistance shows a faster drop and does not follow a simple exponential behaviour suggesting that the dissipation in these two regimes is due to mechanisms of different origin. In the TAFF regime, the vortices are always in motion due to thermal fluctuations in addition to the driving force, and the relaxation time is longer. Fig. 4(c) makes a side-by-side comparison of the behaviour of the resistance at 50 mT (blue) and the time constant shown in Fig. 4(c) as the temperature of the sample is varied from 10 mK to 2 K. We note here that the increase in the resistance and the time-constant against the temperature show a similar behaviour, suggesting the vortex dynamics play a decisive role in establishing the state of the system.

We inspect the NPSD to shine more light on the noise characteristics as we take the system from the BM (B = 50 mT) state to the TAFF state, as shown in Fig. 4(e). We find that throughout the BM regime, until a temperature ~938 mK, the $NP_{int}$ remains unaffected by the temperature. As the system moves into the TAFF regime, we see a progressive rise in the $NP_{int}$ with temperature, suggesting the state is driven by thermal fluctuations. In the BM state, the

$NP_{int}$ influenced by the rise in the field [Fig. 4(b)] and unaffected by temperature variation suggests the pure quantum nature of the state.

The observations in Fig. 3 negate Eddy current heating as a reason for the sudden rise in the resistance. The prominent rise in the resistance is restricted to a narrow ~ 50 mT field window only on the incremental side of the sweep, whereas the Eddy current heating, if significant, should have an effect on the entire field range. The magnetotransport data taken by the stabilized and continuous modes differ from each other only in the vicinity of the secondary minima and maxima, i.e., in a field range of 0-0.6 T in the increasing field direction, corresponding to the BM phase. On highly disordered granular 3D superconducting films, the flux compression can cause a rise in the resistance as the field is increased, but no prominent secondary minima observed[50,51]. The magnetoresistance exhibiting non-monotonous behaviour has been reported on interfacial 2D superconductors and has been attributed to vortex flow, though lacks detailed reasoning or investigation[52].

Though the quest for cleaner disorder-free systems is driven by technology, certain degrees of disorder can be a blessing in disguise. For instance, the disorder enhances the width of resistance plateaus in quantum Hall resistance standards[53]. One needs to strike a delicate balance so that the critical operational parameters, such as coherence times, are not compromised, while maintaining the system dissipation-free. For e.g., vortex qubits need mobile vortices, while their motion[17] will make the superconducting systems dissipative. It has been reported that the choice of measurement environment influences the QM phase and the presence of dissipative circuit elements revives the superconducting state, suggesting the glassy phase as the reason for the origin of the BM phase. The measurement circuit we used consists of low pass filters both at room temperature and also at the 10 mK stage, making the overall bandwidth ~100 Hz. The transition to the BM state at very low fields in our sample does not unambiguously say that the dissipative environment plays a deciding role on the observation of BM state. Owing to the dearth of experiments on clean 2D systems, there is a substantial gap between the experimental observations and the theoretical understanding. We believe that this work has unearthed many aspects of a clean 2D superconducting system most of which are predicted only theoretically.

**Methods**

Few-layered MoS$_2$ flakes are mechanically exfoliated from commercially procured bulk MoS$_2$ (SPI supplies) and transferred onto a highly doped Silicon substrate hosting a 300 nm SiO$_2$ layer using a residue-free dry transfer technique[54]. Electrical contacts to the device is defined by electron-beam lithography followed by Cr (3 nm)/Au (40 nm) metallisation. To minimize any chemical reaction between the ionic liquid and the gold electrodes on the wafer, we restrict the contact of the IL to the sample and the electrodes in the vicinity of the sample using an electron-beam lithography defined PMMA mask. We use DEME-TFSI (N,N-diethyl-N-methyl-N- (2-methoxyethyl) ammonium bis (trifluoromethylsulfonyl)- imide) a popular IL owing to its wide electrochemical window, to electrolytically gate the sample[55]. A lot of care has been exercised to ensure that the fabrication processes are minimally invasive and has been carried out in a low-humidity, clean environment. We refrain from processes such as annealing in reducing atmosphere or plasma cleaning; though these steps have shown to improve contact metal adhesion and contact-resistance, they also introduce defects in the material that could act as vortex pinning centres in the superconducting state. We did not use any capping insulating layer, such as h-BN, as discussed elsewhere[42], instead, the IL was dropped directly on the sample. The capping layer capacitance will add in series with the IL capacitance, requiring the application of a higher IL gate voltages to get the desired level of doping. This can cause an overcrowding of the ions on top of the sample, making the field distribution and doping across the sample non-uniform. After dropping the IL, the samples are quickly mounted onto the mixing chamber of a cryogen-free dilution refrigerator with a base temperature of 10 mK, equipped with an 8 T superconducting magnet, and pumped down to the high-vacuum regime. To minimize chances of any chemical reaction between the sample and the IL, we applied the IL gate voltage only after the sample reached a temperature ~ 250 K, which is above the freezing point of the IL ~ 170 K. The IL gate voltage was ramped at a very slow rate of 1mV/s, providing sufficient time for the ions to rearrange on the sample uniformly. All the transport data are acquired with a combination of low-noise, low-frequency lock-in (SRS 830) and low-noise DC measurements employing a low-noise voltage preamplifier (SRS 560). Magneto-transport data are acquired using an 8 T superconducting magnet in two different modes; (i) stabilized: by stabilizing the electrical behaviour of the material for several minutes ~20 min. or, (ii) continuous: by continuously sweeping the magnetic field at various rates to capture the transient behaviour. The sweep rates are kept low to avoid any Eddy-current heating effects. For this, the data from the sample has been acquired after a wait time of 1s once the desired magnetic field has been reached.

**Author contributions**

MT conceived the problem. SN & AK prepared all the samples. SN, AK & AAS conducted the transport measurements. SN, AK & MT analysed the data. SN & MT co-wrote the manuscript.

**Acknowledgements**

MT acknowledge SERB, Govt. of India for the financial support under the grand CRG/2018/004213, Chithra H Sharma, Siddhartha Lal & Radhikesh Raveendran for the scientific discussions.

# Transient vortex dynamics and evolution of Bose metal from a 2D superconductor on MoS$_2$

Sreevidya Narayanan[#], Anoop Kamalasanan[#], Annu Anns Sunny and Madhu Thalakulam[1]

Indian Institute of Science Education & Research Thiruvananthapuram, Kerala 695551, India

**S1: Hall measurement data:** We extract a carrier concentration of $9.2\times10^{14} \text{cm}^{-2}$ from the low-field Hall measurement data taken at 4 K which is higher than $6\times10^{13} \text{cm}^{-2}$ observed to obtain 2D superconductivity in MoS$_2$.

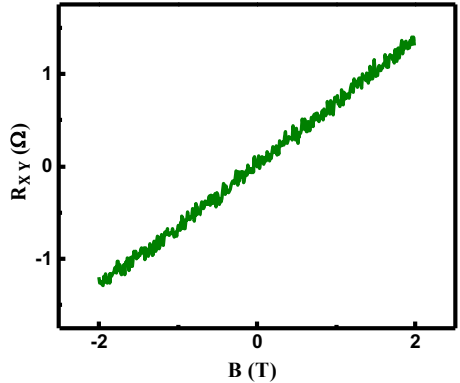

Transverse resistance ($R_{XY}$) versus Magnetic field at 4 K

**S2: Ioffe-Regal criteria:** The Ioffle – Regel criteria for metallic conduction requires $K_F.l_e \geq 1$ where $K_F = \sqrt{2\pi n_{2D}}$ is the Fermi wave vector. With $l_e = 67\ nm$ and $n_{2D} = 9.2\times10^{14} cm^{-2}$ for our device, $K_F.l_e \sim 552$, well exceed the criteria, validating the metallic nature of our device.

**S3: Quantum creep model:** For quantum creep model, the resistance assumes an exponential dependence on the applied field given by

$R = h/4e^2 exp[C\pi/2(\frac{\frac{h}{2\pi e^2}}{R_N})(B - B_{C2})/B_{C2}]$. Data shows a poor fit to the quantum creep model.

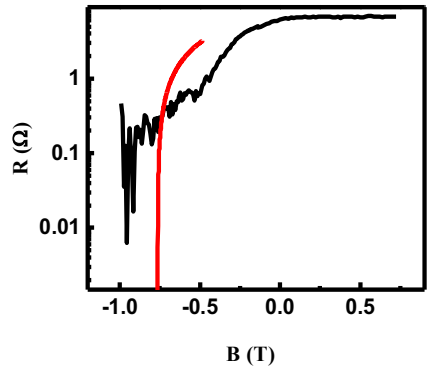

Fit (red) using Quantum creep model

**S4: Transient behaviour of magnetoresistance**

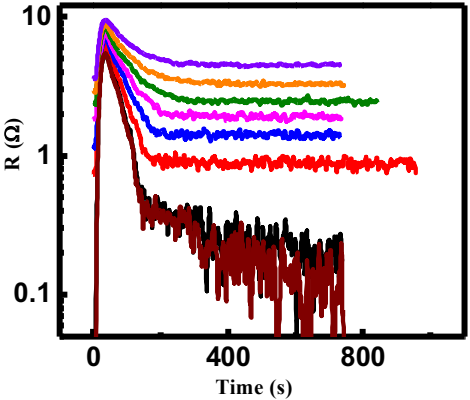

Transient behaviour of magnetoresistance for temperatures 10 mK (brown), 514 mK (black), 935 mK (red), 1.15 K (blue), 1.25 K (magenta), 1.45 K (green), 1.56 K (orange), 1.77 K (violet) for the field step 0 T to -0.05 T.